\documentclass[twocolumn,aps,prl,superscriptaddress]{revtex4}

\usepackage{amssymb}

\makeatletter

\begin{document}

\title{Resilient Quantum Computation in Correlated Environments: \\
A Quantum Phase Transition Perspective}

\author{E. Novais} 

\affiliation{Department of Physics, Duke University,
Durham, North Carolina 27708-0305}

\author{Eduardo R. Mucciolo}

\affiliation{Department of Physics, University of Central Florida,
Orlando, Florida 32816-2385}

\author{Harold U. Baranger}

\affiliation{Department of Physics, Duke University, Durham, North
Carolina 27708-0305}

\date{\today}

\begin{abstract}
We analyze the problem of a quantum computer in a correlated
environment protected from decoherence by QEC using a perturbative
renormalization group approach. The scaling equation obtained reflects
the competition between the dimension of the computer and the scaling
dimension of the correlations. For an irrelevant flow, the error
probability is reduced to a stochastic form for long time and/or large
number of qubits; thus, the traditional derivation of the threshold
theorem holds for these error models. In this way, the ``threshold
theorem'' of quantum computing is rephrased as a dimensional
criterion.
\end{abstract}

\maketitle

A central result in the theory of quantum error correction (QEC) is
the ``threshold theorem'' \cite{NC00}. Even though QEC is a
perturbative method \cite{Ste96+CS96,NC00}, the ``threshold theorem''
states that: Provided the noise strength is below a critical value,
quantum information can be protected for arbitrarily long times. This
remarkable theorem was first derived for stochastic error models
\cite{KLZ98b+KLZ01} or, more generally, for environments where the
spatial and temporal correlations have an exponential decay
\cite{DAMB99}. Although there is still some controversy \cite{ALZ06},
the theorem is widely accepted by a community where the foremost
effort has been to extend it to correlated error models
\cite{TB05,AGP06+Rei05,KLV00,AKP06}.

The key difference between a correlated error model and a stochastic
one is the absence of a local and time independent error probability.
The main concept in extending the theorem to a correlated environment
has been the norm of an operator \cite{KLV00,TB05,AKP06}. A convenient
norm of the interaction Hamiltonian can be used to bound the error
probability and, eventually, to prove the threshold theorem. However,
for some models the norm of the interacting Hamiltonian is too large
to provide useful bounds \cite{TB05}. In this paper, we hence choose
to follow different reasoning.

We demonstrate that a large class of correlated error models is
reduced to a simple stochastic error model in the asymptotic limit of
large number of qubits or long time. Thus, in order to prove the
resilience of the quantum information, we can fall back on the
traditional derivation of the threshold theorem. Because the
conditions for this fall back have clear parallels with the theory of
quantum phase transitions \cite{NO98,Sac99}, we rephrase the threshold
theorem as a dimensional criterion: (i) For systems above their
``critical dimension'', the traditional proof of resilience is valid,
and there are two regimes, or phases, as a function of the coupling
with the environment. (ii) However, when the system is below its
``critical dimension'', the effects of correlations produce large
corrections, and it is not possible to prove resilience by our
arguments.

In an elegant paper, Aharonov, Kitaev, and Preskill \cite{AKP06} (AKP)
derived a new set of ``threshold conditions'' for a correlated noise
model with instantaneous power law interactions between any two qubits
of the computer. AKP proved that a version of the theorem holds if:
(i) the coupling is below a critical value, and (ii) the interaction
between the qubits decays sufficiently fast when compared with the
number of spatial dimensions of the computer. The first condition is
very much what one expects for the ``threshold theorem''. However, the
second is very suggestive of our interpretation of the threshold
theorem in terms of a quantum phase transition. Even though AKP
considered an error model substantially different from ours (see
below), a particular case of our discussion reproduces AKP's
conditions for resilience.

\emph{The error model---} We consider an environment described by a
``non-interacting'' field theory, $H_{0}$, allowing application of
Wick's theorem. Furthermore, since an environment with a spectral gap
would imply exponentially decaying correlations \cite{DAMB99}, we
assume that the environment is gapless, has wave velocity $v$, and
short time cutoff $\Lambda^{-1}$. We focus on an interaction
Hamiltonian that is local in the qubits,
\begin{equation}
V = \sum_{{\bf x},\alpha}\lambda_{\alpha}f_{\alpha}\left({\bf {\bf
x}}\right)\sigma_{\alpha}\left({\bf {\bf x}}\right),
\label{eq:int-hamil}
\end{equation}
where $\vec{f}$ is a function of the environment degrees of freedom
and $\vec{\sigma}$ are the Pauli matrices that parametrize the
qubits. Finally, we consider (without loss of generality) the qubits
to be arranged in a $D$-dimensional lattice.

The evolution of the system in the interaction picture during a QEC
cycle is given by
\begin{eqnarray}
\hat{U}\left(\Delta,\lambda_{\alpha}\right) & = &
T_{t} \, e^{-i\int_{0}^{\Delta}dt \sum_{{\bf x},\alpha} \lambda_{\alpha}
f_{\alpha}({\bf x},t) \sigma_{\alpha}({\bf x}) } \;,
\label{eq:full-evol}
\end{eqnarray}
with $\Delta$ corresponding to the time of the syndrome extraction,
$T_{t}$ the time ordering operator, and
$f_{\alpha}({\bf x},t) =e^{\frac{i}{\hbar}H_{0}t} f_{\alpha}({\bf x})
e^{-\frac{i}{\hbar}H_{0}t}$.  At time $\Delta$, the extraction of
syndromes selects some terms of Eq.\,(\ref{eq:full-evol}) as the
evolution operator for that particular QEC cycle \cite{NB05}. Hence,
we naturally define a coarse-grained space-time grid of hypercubes,
$\Delta\times\left(v\Delta\right)^{D/z}$
(with $z$ being the dynamical exponent of the environment).

The coarse grain grid is the frame upon which we develop our analysis.
It provides the scale that separates two distinct noise regimes:
intra- and inter-hypercube components. The key simplifying assumption
is that in each volume $\left(v\Delta\right)^{D/z}$ there is only
\textit{one} qubit. In this case, the intra-hypercube part is simply
the probability $\epsilon_{\alpha}$ of an error of type
$\sigma_{\alpha}$ on that particular qubit.
Two physical systems where this hypothesis should be immediately valid
are acoustic phonons interacting with solid-state qubits and
ohmic noise due to voltage/current fluctuations on qubits based
on quantum dots or superconducting devices.

Our discussion is
therefore divided into two parts. First, we demonstrate how to
calculate this ``stochastic'' error probability, $\epsilon_{\alpha}$.
Second, we evaluate how this error probability is changed by the
inter-hypercube component of the noise.

\emph{Defining $\epsilon_{\alpha}$---} Consider, for instance, that we
know from the syndrome that a particular qubit, say ${\bf x_{1}}$,
suffered an error $\sigma_{\alpha}$ in the time interval labeled
$0$. To lowest order in $\lambda_\alpha$, the QEC code disentangles
the qubit from the environment. Thus,
\begin{equation}
\hat{\upsilon}_{\alpha}\left({\bf x}_{1},\lambda_{\alpha}\right)
\approx -i\lambda_{\alpha}\int_{0}^{\Delta}dt\, f_{\alpha}\left({\bf
{\bf x}_{1}},t\right),\label{eq:lowestorder-evol}
\end{equation}
is the evolution operator associated with that particular qubit, where
we used that $\sigma_{\alpha} \left(\Delta\right) \sigma_{\alpha}
\left(t\right) = {\bf 1}$ \cite{footnote1}.

Of course, lowest order perturbation theory is not always
justified. Hence, before QEC can be argued to be effective, it is
important to estimate how higher order terms change
Eq.\,(\ref{eq:lowestorder-evol}). A very direct approach is to use the
perturbative renormalization group (RG).

To derive the RG equations, we supplement
(\ref{eq:lowestorder-evol}) with the next higher-order terms
allowed by the QEC code
\begin{widetext}
\begin{eqnarray}
\hat{\upsilon}_{\alpha}\left({\bf x}_{1},\lambda_{\alpha}\right) &
\approx & -i\lambda_{\alpha} \int_{0}^{\Delta}dt\,
f_{\alpha}\left({\bf x_{1}},t\right) - \frac{1}{2}
\left|\epsilon_{\alpha\beta\gamma}\right| \lambda_{\beta}
\lambda_{\gamma}\, \sigma_{\alpha} \left(\Delta\right) T_{t}
\int_{0}^{\Delta} dt_{1}\, dt_{2}\, f_{\beta} \left( {\bf x_{1}},t_{1}
\right) f_{\gamma} \left( {\bf x_{1}},t_{2} \right) \sigma_{\beta}
\left( t_{1} \right) \sigma_{\gamma} \left( t_{2} \right) \nonumber \\
& + & \frac{i}{6} \sum_\beta \lambda_{\alpha} \lambda_{\beta}^{2}\,
\sigma_{\alpha} \left( \Delta \right) T_{t} \int_{0}^{\Delta} dt_{1}\,
dt_{2}\, dt_{3}\, f_{\alpha} \left( {\bf x_{1}},t_{1} \right)
f_{\beta} \left( {\bf x_{1}},t_{2} \right) f_{\beta} \left( {\bf
x_{1}},t_{3}\right)\sigma_{\alpha}\left( t_{1} \right) \sigma_{\beta}
\left( t_{2} \right) \sigma_{\beta} \left( t_{3}
\right),\label{eq:extended-evolution}
\end{eqnarray}
\end{widetext}
where $\epsilon_{\alpha\beta\gamma}$ is the antisymmetric
tensor. \textit{Since QEC completely removes the qubit variable from
the problem, we can consider Eq.\,(\ref{eq:extended-evolution}) as a
field theory problem in itself.}
It is therefore straightforward to
derive the lowest order terms to the $\beta$-function of $\hat{\upsilon}_\alpha ({\bf x_{1}})$,
\begin{equation}
\frac{d\lambda_{\alpha}}{d\ell} = g_{\beta\gamma} \left(\ell\right)
\lambda_{\beta} \lambda_{\gamma} + \sum_{\beta}h_{\alpha\beta}
\left(\ell\right) \lambda_{\alpha} \lambda_{\beta}^{2},\label{betafunction}
\end{equation}
where $g$ and $h$ are functions specific to a particular environment
and $d\ell= d\Lambda/\Lambda$ \cite{Affleck}. Note that the integration of
Eq.~(\ref{betafunction})
is equivalent to summing an infinite series of terms in the perturbative expansion
for $\hat{\upsilon}_\alpha ({\bf x_{1}})$.
Because there is only one qubit in the hypercube, the problem of
calculating $\epsilon_{\alpha}$ involves only $f({\bf x_1})$ and so was
reduced to an impurity problem. 

To further proceed with the argument, we must ask that the
renormalized value of $\lambda_{\alpha}$ at the
$\left(v\Delta\right)^{-1}$scale, $\lambda_{\alpha}^{*}$, be a small
parameter. In that case, it is appropriate to use the evolution
Eq.\,(\ref{eq:lowestorder-evol}) with $\lambda_{\alpha}$ replaced by
$\lambda_{\alpha}^{*}$.

The conditional probability of having an error of type $\alpha$ in a
particular hypercube labeled by ${\bf x_1}$ and $t=0$ has the general
form \cite{NB05}
\begin{eqnarray*}
P\left(...;\alpha,{\bf x}_{1};...\right) & \approx & \left\langle
...\hat{\upsilon}_{\alpha}^{\dagger}\left({\bf
x}_{1},\lambda_{\alpha}^{*}\right)...\hat{\upsilon}_{\alpha}\left({\bf
x}_{1},\lambda_{\alpha}^{*}\right)...\right\rangle .
\end{eqnarray*}
Hence, following the discussion in Ref. \cite{NB05}, we define the
operator that gives the probability of an error as
\begin{equation}
\upsilon_{\alpha}^{2}\left({\bf x}_{1},\lambda_{\alpha}^{*}\right)
\approx \hat{\upsilon}_{\alpha}^{\dagger}\left({\bf
x}_{1},\lambda_{\alpha}^{*}\right)\hat{\upsilon}_{\alpha}\left({\bf
x}_{1},\lambda_{\alpha}^{*}\right).
\end{equation}
We can now readily separate the effects of correlations into their
intra- and inter-hypercube parts. Using Wick's theorem, we obtain
\begin{equation}
\upsilon_{\alpha}^{2} \left( {\bf x}_{1},\lambda_{\alpha}^{*} \right)
\approx \epsilon_{\alpha} + \left( \lambda_{\alpha}^{*} \Delta
\right)^{2} \,:\! \left| f_{\alpha} \left( {\bf x_{1}},0 \right)
\right|^{2} \!: \;,
\label{eq:error-prob}
\end{equation}
where the intra-hypercube part is
\begin{equation}
\epsilon_{\alpha} = \left( \lambda_{\alpha}^{*} \right)^{2}
\int_{0}^{\Delta} dt_{1} \int_{0}^{\Delta} dt_{2}\,\left\langle
f_{\alpha}^{\dagger} \left( {\bf x_{1}},t_{1} \right) f_{\alpha}
\left( {\bf x_{1}},t_{2} \right) \right\rangle
\end{equation}
and $::$ stands for normal ordering. An important remark is that
$\epsilon_{\alpha}$ is from this point on a numerical factor set by
$\lambda^{*}$ at the scale $\left(v\Delta\right)^{-1}$.

For later convenience, we re-write Eq.\,(\ref{eq:error-prob}) in a
slightly different form. Defining the operators
\begin{eqnarray}
F_{\alpha}\left({\bf x_{1}},0\right) & = &
\frac{\left(\lambda_{\alpha}^{*}\Delta\right)^{2}}
{\epsilon_{\alpha}} \,:\! \left|f_{\alpha}\left({\bf
x_{1}},0\right)\right|^{2} \!: \;,\label{eq:Fa}
\end{eqnarray}
we re-write the operator for the probability of an error as the product
$\upsilon_{\alpha}^{2}\left({\bf x_{1}},\lambda_{\alpha}^{*}\right) =
\epsilon_{\alpha}\left[1+F_{\alpha}\left({\bf x_{1}},0\right)\right]$.
By direct calculation, or simply by the unitarity of the probability,
it is also straightforward to write the operator for the probability
of not having an error as $\upsilon_{0}^{2}\left({\bf
x_{1}},\lambda_{\alpha}^{*}\right)=$
$\left[1-\sum_{\alpha}\epsilon_{\alpha}\right]F_{0}\left({\bf
x_{1}},0\right)$, where
\begin{eqnarray}
F_{0}\left({\bf x_{1}},0\right) & = & 1 - \frac{\sum_{\beta} \left(
\lambda_{\beta}^{*} \Delta \right)^{2} \,:\! \left| f_{\beta} \left( {\bf
x_{1}},0 \right) \right|^{2}\!:} {1 - \sum_{\beta=x,y,z}
\epsilon_{\beta}}.
\label{eq:F0}
\end{eqnarray}

\emph{Probability function and scaling---} Now that we separated the
probability into an intra- and inter-hypercube component, we can seek
to write the probability of a computer evolution with $m$ errors after
$N$ QEC cycles. A particular case is helpful in understanding how to
proceed. Suppose we want to calculate the probability that $R$ qubits
suffer $m\ll NR$ errors of type $\alpha$. Using
Eqs. (\ref{eq:error-prob})-(\ref{eq:Fa}), it is straightforward to
write\begin{widetext}
\begin{equation}
P_{m}^{\alpha} = p_{m}\int\frac{d{\bf
x}_{1}}{\left(v\Delta\right)^{D/z}}...\frac{d{\bf
x}_{m}}{\left(v\Delta\right)^{D/z}}
\int_{0}^{N\Delta}\frac{dt_{1}}{\Delta}...\int_{0}^{t_{m-1}}
\frac{dt_{m}}{\Delta} \left\langle \Big[ \prod_{\zeta}F_{0}({\bf
x_{\zeta}},t_{\zeta})\Big] \big[1+F_{\alpha}({\bf x_1},t_1)\big]...
\big[1+F_{\alpha}({\bf x_m},t_m)\big] \right\rangle \label{eq:propm}
\end{equation}
\end{widetext}
where we integrated over all possible grid positions, $\left({\bf
x}_{j},t_{j}\right)$, $\zeta$ denotes the set of remaining hypercubes,
and $p_{m} = \left(1-\sum_{\alpha}\epsilon_{\alpha}\right)^{RN-m}
\left(\epsilon_{\alpha}\right)^{m}$ \cite{footnote2}.

We now organize the expectation value of Eq.\,(\ref{eq:propm}) in
powers of $\left(\lambda^{*},\epsilon_{\alpha}\right)$ and invoke
Wick's theorem again. The first term is just the stochastic
contribution to the probability,
\begin{equation}
p_{m}\int\prod_{k=1}^{m}\frac{d{\bf {\bf
x}}_{k}}{\left(v\Delta\right)^{D/z}}\frac{dt_{k}}{\Delta} =
p_{m}\left(\!\!\begin{array}{c} NR\\ m\end{array}\!\right)\sim
p_{m}\left(NR\right)^{m}.\label{eq:stoch}
\end{equation}
The next term is typically of the form 
\begin{equation}
p_{m}\int\prod_{k=1}^{m}\frac{d{\bf {\bf
x}}_{k}}{\left(v\Delta\right)^{D/z}}\frac{dt_{k}}{\Delta}\left\langle
F_{\alpha}\left({\bf x_{i}},t_{i}\right)F_{\alpha}\left({\bf
x_{j}},t_{j}\right)\right\rangle .\label{eq:firstcorrection}
\end{equation}
Thus, the fundamental role of the scaling dimension of $F_{\alpha}$
now becomes clear.  If $\dim\big[f_{\alpha}\big] = \delta_{\alpha}$,
then Eq.\,(\ref{eq:Fa}) implies that $\dim\big[F_{\alpha}\big] =
2\delta_{\alpha}$.  Since $H_{0}$ is assumed to be non-interacting,
the two point correlation function has the general form
$$ \left\langle F_{\alpha} \left( {\bf {\bf x}_{i}},t_{i} \right)
F_{\alpha}\left({\bf x_{j}},t_{j}\right)\right\rangle \sim \mathcal{F}
\left( \left| {\bf x}_{i} - {\bf x}_{j} \right|^{-4\delta_{\alpha}}\!,
\left|t_{i}-t_{j} \right|^{-4\delta_{\alpha}/z}\right)
$$
For large $N$ or $R$, we can study the stability of the expansion of Eq.~(\ref{eq:propm})
in powers of $\lambda^{*}$ using the traditional scaling theory \cite{NO98}.
The simplest way to proceed is to apply the  transformation
${\bf x_{i,j}} \to b{\bf x_{i,j}^{\prime}}$ and $t_{i,j} \to b^{z} t_{i,j} ^{\prime}$,
with $b=e^{d\ell}$, to  Eq.~(\ref{eq:firstcorrection}). It is then straightforward to obtain the \textit{scaling equation for $\lambda^{*}$,}
\begin{eqnarray}
\frac{d\lambda_{\alpha}^{*}}{d\ell} & = &
\left(D+z-\dim\left[F_{\alpha}\right]\right)\lambda_{\alpha}^{*}.
\label{eq:dimensionequation}
\end{eqnarray}
A similar argument can be applied to all the remaining terms of
Eq.\,(\ref{eq:propm}). The critical condition is when $D+z = 2\delta$.
In this particular case, correlations between hypercubes introduce logarithmic
corrections to the stochastic part of Eq.~(\ref{eq:propm}).

An irrelevant flow for $\lambda^{*}$ indicates that the system is
above its ``critical dimension''. Since correlations between
hypercubes produce small corrections in comparison to 
Eq.\,(\ref{eq:stoch}), the probability distribution at long time and/or
large number of qubits has essentially a stochastic form. Corrections
to this form can be systematically calculated by perturbative
expansion in $\lambda_{\alpha}^{*}$. Thus, we can fall back on the
usual proof of resilience for quantum information
\cite{KLZ98b+KLZ01,DAMB99}: There will be two ``phases'' separated by
a $\lambda_{\alpha}^{\rm critical}$: (i) For
$\lambda_{\alpha}<\lambda_{\alpha}^{\rm critical}$, the information
can be protected for arbitrarily long times. Hence, the computer and
the environment are disentangled by the QEC code. (ii) In contrast,
for $\lambda_{\alpha}>\lambda_{\alpha}^{\rm critical}$, the computer
and the environment can not be disentangled and decoherence will take
place.

This scenario strongly resembles the theory of quantum phase
transitions. There are several ways to pursue this analogy; here we
present two. First, we just showed that for systems above their
critical dimension, the second term of the r.h.s. of
Eq.\,(\ref{eq:error-prob}) is unimportant at asymptotically large
scales. In other words, when we calculate the probability
$P\left(....;\alpha,{\bf x}_{1};...\right)$, it is a good
approximation to replace the operator $v_{\alpha}^{2}\left({\bf
x}_{1},\lambda_{\alpha}^{*}\right)$ by its ``mean field''
(perturbative) value $\epsilon_{\alpha}$. The bare value of
$\lambda_{\alpha}$ is the only parameter that determines whether QEC
can succeed and thus plays a role analogous to temperature. This is
precisely what one expects for systems above their upper critical
dimension, where the transition has a mean field character
\cite{NO98}.

Second, an explicit analogy with a statistical mechanical problem is
also possible. Eq.\,(\ref{eq:propm}) can be thought of as the partition
function of a gas of fictitious particles, where
$\lambda_{\alpha}^{*}$ is the particle fugacity, $F_{\alpha}$ creates
a particle, and $F_{0}$ introduces the vacuum fluctuations. An example
is quite illuminating: Consider the simple case of a bosonic bath with
$(\phi,\pi)$ representing the conjugate fields. For a $D=1$ computer
and a noise model $V=\lambda\sum_{j}\cos\left[\delta \cdot
\phi\left(j\right)\right]\sigma_{j}^{z}$, Eq.\,(\ref{eq:propm}) is
precisely the partition function of a 2-dimensional Coulomb gas. This
case, then, has a quantum phase transition in the Kosterlitz-Thouless
universality class \cite{Niehnus} as a function of $\delta$.

On the other hand, going back to Eq.\,(\ref{eq:dimensionequation}), a
relevant flow for $\lambda^{*}$ indicates that inter-hypercube
correlations produce contributions to the probability function that
scale in the same fashion as the intra-hypercube part.  In this sense,
there is no ``de facto'' separation of scales or, in other words, the
problem is inherently non-perturbative in the coupling with the
environment. It is therefore not possible to prove resilience by our
arguments.

\emph{Changing $\dim\left[F_{\alpha}\right]$---} We showed previously
\cite{NB05} that small changes in the QEC code can dramatically reduce
the effects of correlations between the hypercubes. By simply applying
logical NOTs and phase-NOTs in each QEC cycle, the dimension of
$F_{\alpha}$ becomes larger at the cost of increasing
$\epsilon_{\alpha}$. To understand this fact, revisit
Eq.\,(\ref{eq:lowestorder-evol}); for illustration, consider $\alpha=z$
and apply a logical NOT half way through the cycle:
\begin{eqnarray*}
\hat{\upsilon}_{z}\left({\bf x}_{1},\lambda\right) & \approx &
i\lambda_{z} \left[ \int_{\Delta/2}^{\Delta} dt - \int_{0}^{\Delta/2}
dt\right]\, f_{z}\left({\bf {\bf x}_{1}},t\right).
\end{eqnarray*}
Following the same steps as before, we obtain that $F_{z}\left({\bf
x_{1}},0\right)\propto \,\,:\!\! |\partial_{t}f_{z}\left({\bf
x_{1}},0\right)|^{2} \!\!: $. If $n$ logical NOTs and phase-NOTs are 
performed, we find that $\dim\left[F_{\alpha}\right] = 
2\left(\delta_{\alpha}+nz\right)$. \textit{Therefore, for a given noise 
model, one can always engineer an irrelevant flow.}

\emph{Connection to the AKP results---} In Ref. \cite{AKP06} a family
of long-ranged noise models with interactions between qubits was
studied. AKP considered a power law interaction between any two qubits
at positions ${\bf x}_{1}$ and ${\bf x}_{2}$ of the computer with
strength $\Gamma|{\bf x}_{1}- {\bf x}_{2}|^{-2\delta}$. Clearly, 
one could start from their noise
model and use a Hubbard-Stratonovich transformation to arrive at
ours. The reverse is also true: starting from our model, one could
integrate out the environment field and arrive at the effective
interaction between qubits that AKP considered.

Nevertheless, the two papers deal with opposite limits of this same
model. The crucial difference is the wave velocity of the environment,
$v$. AKP considered the limit of instantaneous interactions,
$v\to\infty$. This implies that \textit{all} the qubits of the
computer are contained in the volume $\left(v\Delta\right)^{D/z}$, in
contrast to our assumption that \textit{one} qubit is in that
volume. Interestingly, the AKP limit leads to an important
simplification when combined with QEC: Since errors are detected by
QEC, there are no memory effects due to correlations between qubits at
different QEC steps. Hence, their error model corresponds in our
analysis to a dynamical exponent $z=0$. With this, our criterion for
the possibility of resiliency is exactly the same as AKP's, even
though the problem and methods used are inherently different. We
speculate, therefore, that there is a more general scheme that
encompasses both papers.

\emph{Remarks and conclusions---} In hindsight, the results that we
obtain are dramatically clear and simple. QEC is a perturbative
method. Therefore, the ``threshold condition'' should be a statement
about when a perturbative analysis is valid. In field theories this is
a relatively straightforward question. However, in a quantum computer
this is not so obvious due to the presence of the qubits. The
remarkable result of QEC is that some of the quantities relevant for
the threshold condition depend exclusively on the environment
variables. It is, then, possible to derive criteria based only on the
field theory that describes the environment.

Our discussion brings to light an interesting parallel with the theory
of quantum phase transitions. In fact, we derived a condition that
strongly resembles the definition of the ``upper critical dimension''
of a quantum problem. When the system is above this dimension, the
usual ``mean field'' derivation of the threshold theorem is
applicable. Hence, as a function of the bare coupling with the
environment, there is a transition from the states of the qubits and
the environment being disentangled to their being entangled.

An important question that remains open is if there is also a ``lower
critical dimension'', namely a criterion for the impossibility of
proving resilience. If the lower and upper critical dimensions are not
the same, the intermediate cases would require a proof of resilience
substantially different from the ``mean field'' approach used here. An
example of such an approach is the use of the norm of the interaction
Hamiltonian by AKP. However, the fact that they found the same
``critical dimension'' as we have found suggests that the ``upper''
and ``lower'' critical dimensions may coincide.

% In conclusion, we analyzed the problem of a quantum computer in a
% correlated environment protected from decoherence by QEC. We discussed
% whenever resilience of the quantum information can be achieved in
% correlated environments. Finally, we presented an interesting analogy
% with the theory of quantum phase transitions. 

We thank C. Kane and D. Khveshchenko for helpful discussions.
This work was supported in part by NSF Grants No. CCF 0523509 and
0523603. ERM acknowledges partial support from the I$^2$Lab at
UCF. ERM and HUB thank the Aspen Center for Physics for its
hospitality.

%\vspace{-.5cm}

%\bibliographystyle{apsrev}
%\bibliography{newletter-bib,footnotes}

\end{document}